\documentclass[%
prx,
amsmath,amssymb,
reprint,%
superscriptaddress,
]{revtex4-2}

\usepackage{graphicx}
\usepackage{dcolumn}
\usepackage{bm}
\usepackage{tabularx}
\usepackage{graphicx}
\usepackage{multirow}
\usepackage{siunitx,color}
\usepackage{booktabs}
\usepackage[colorlinks=true,citecolor=blue
]{hyperref}
\usepackage[version=3]{mhchem}

\begin{document}
	\title{Tunable Electronic Properties and Band Alignments of \ce{MoSi2N4}/GaN and \ce{MoSi2N4}/ZnO van der Waals Heterostructures}
	\author{Jin Quan Ng}
	\affiliation{Science, Mathematics and Technology,
		Singapore University of Technology and Design (SUTD), 8 Somapah Road,
		Singapore 487372, Singapore}	
	
	\author{Qingyun Wu}
	\affiliation{Science, Mathematics and Technology,
		Singapore University of Technology and Design (SUTD), 8 Somapah Road,
		Singapore 487372, Singapore}
	
	\author{L. K. Ang}
	\thanks{Authors to whom correspondence should be addressed: ricky\_ang@sutd.edu.sg and yeesin\_ang@sutd.edu.sg}
	\affiliation{Science, Mathematics and Technology,
		Singapore University of Technology and Design (SUTD), 8 Somapah Road,
		Singapore 487372, Singapore}
	
	\author{Yee Sin Ang}
	\thanks{Authors to whom correspondence should be addressed: ricky\_ang@sutd.edu.sg and yeesin\_ang@sutd.edu.sg}
	\affiliation{Science, Mathematics and Technology,
		Singapore University of Technology and Design (SUTD), 8 Somapah Road,
		Singapore 487372, Singapore}
	
	\begin{abstract}
		Van de Waals heterostructures (VDWH) is an emerging strategy to engineer the electronic properties of two-dimensional (2D) material systems. Motivated by the recent discovery of MoSi$_2$N$_4$ - a synthetic septuple-layered 2D semiconductor with exceptional mechanical and electronic properties, we investigate the synergy of \ce{MoSi2N4} with wide band gap (WBG) 2D monolayers of GaN and ZnO using first-principle calculations. We find that \ce{MoSi2N4}/\ce{GaN} is a direct band gap Type-I VDWH while \ce{MoSi2N4}/\ce{ZnO} is an indirect band gap Type-II VDWH.
		Intriguingly, by applying an electric field or mechanical strain along the out-of-plane direction, the band structures of \ce{MoSi2N4}/\ce{GaN} and \ce{MoSi2N4}/\ce{ZnO} can be substantially modified, exhibiting rich transitional behaviors, such as the Type-I-to-Type-II band alignment and the direct-to-indirect band gap transitions. These findings reveal the potentials of \ce{MoSi2N4}-based WBG VDWH as a tunable hybrid materials with enormous design flexibility in ultracompact optoelectronic applications.
	\end{abstract}
	\maketitle
	
	In recent years, van der Waals heterostructures (VDWHs) have been widely employed for engineering the electronic, optical and photocatalytic properties of two-dimensional (2D) materials \cite{liu2016van}. 
	With an appropriate selection of 2D monolayers and stacking order, 
	VDWHs can give rise to myriads of interesting physics, such as strongly interacting artificial heavy fermions \cite{vavno2021artificial}, and excitonic Bose-Einstein Condensates \cite{liu2017quantum}. 
	VDWHs also offer an avenue to engineer high-quality electrical contacts \cite{cao2021two, cao2020electrical} with exceptionally low contact resistance and significantly suppressed Fermi level pinning effect, thus playing a pivotal role towards the development of practical 2D semiconductor device technology \cite{liu2018approaching, liu2016vanweakfermi}.
	
	As such, VDWHs have the ability to enable applications in new devices not available with currently available materials or enhance current devices with new characteristics. One example of which are tunnelling transistors which operate by tunnelling current \cite{hofstein1965insulated}. Such devices made using VDWHs promise to have subthreshold swings below traditional MOSFET limits and hence have lower off currents and standby power dissipation, thus better suited for low power applications \cite{koswatta2009performance}. Another example are better photodetectors with fast switching speed \cite{gan2013chip} due to narrow channel width, higher responsivity due to built-in electric fields and broad spectral bandwidth \cite{yu2017near}.
	
	\begin{figure*}[t]
		\centering
		\includegraphics[width=0.95\textwidth]{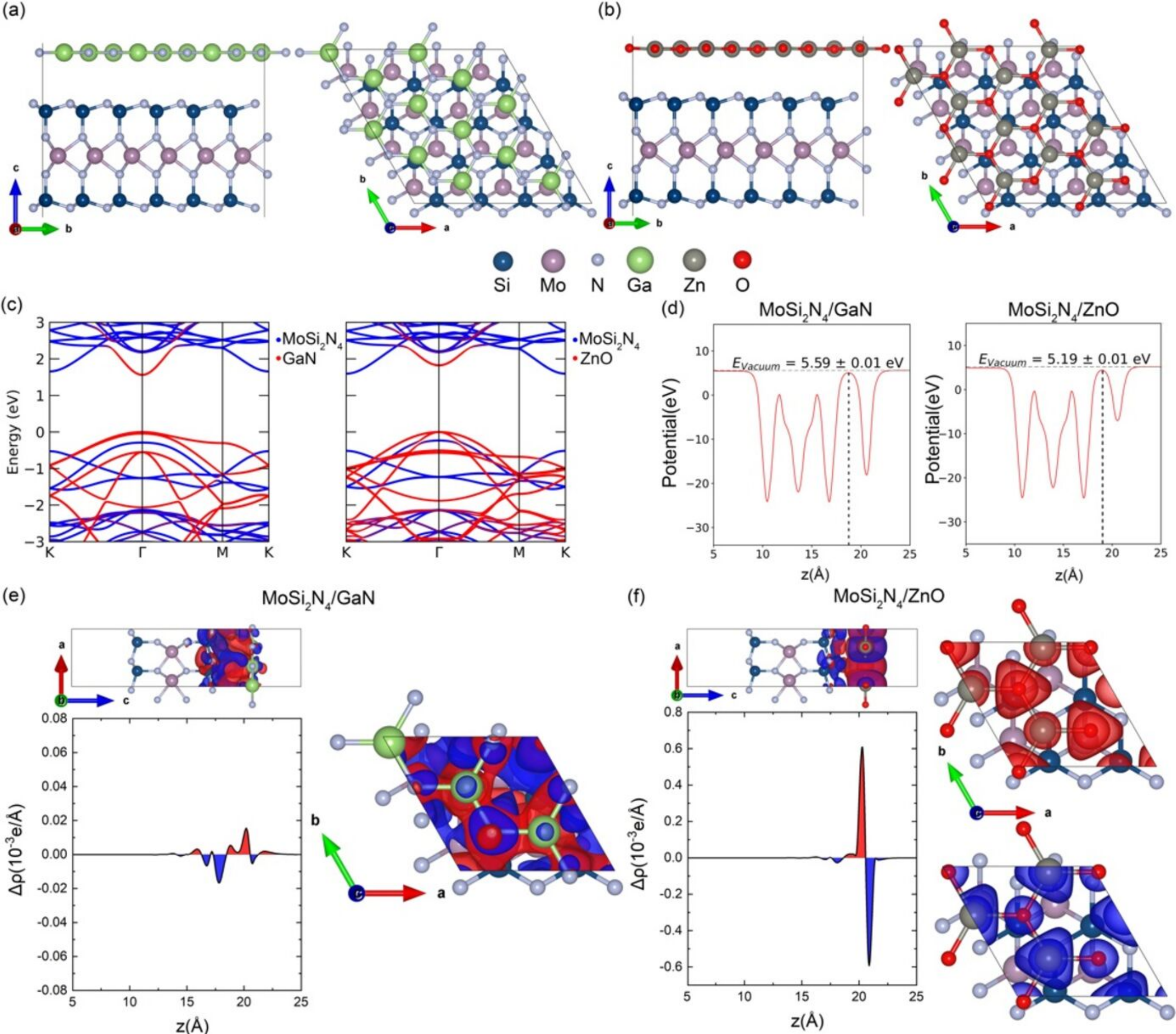}
		\caption{(a) Top and side view of \ce{MoSi2N4}/GaN (b) Top and side view of \ce{MoSi2N4}/ZnO. (c) Band structures of \ce{MoSi2N4}/GaN and \ce{MoSi2N4}/ZnO.(d) Potential plots of \ce{MoSi2N4}/GaN and \ce{MoSi2N4}/ZnO, with the black dotted line showing heterostructure interface. (e) Left hand side, charge density difference plots showing cross section isosurface and charge density difference in the of \ce{MoSi2N4}/GaN, red and blue showing charge accumulation and depletion respectively. Right hand side, isosurface of charge density difference from the top (f)  Left hand side, charge density difference plots showing cross section isosurface and charge density difference of \ce{MoSi2N4}/ZnO. Right hand side, isosurface of charge density difference from the top}%
		\label{Figure1}%
	\end{figure*}
	
	The recent discovery of \ce{MoSi2N4} and the extended \ce{MA2Z4} monolayer family reveals an exciting material platform 
	for designing novel 2D material devices. \ce{MoSi2N4} monolayer is a \textit{synthetic} 2D semiconductor without 
	3D parent structure and has been synthesized experimentally by passivating \ce{MoN2} monolayer with Si to create a 
	septuple-layered nanosheet composed of a \ce{MoN2} inner layer sandwiched by two Si-N outer layers \cite{hong2020chemical}. 
	Interestingly, the Conduction Band Minimum (CBM) and the Valence Band Maximum (VBM) are concentrated within the \ce{MoN2} core-layer
	and the outer Si-N atomic sublayer, hence serving as a built-in protective layer to preserve the conduction channels from the
	external perturbations \cite{wang2021efficient, liu2018approaching} and to strongly suppress the adverse Fermi level pinning effect. 
	Quantum transport simulations have also suggested \ce{MoSi2N4} monolayer to be an exceptional 2D channel material 
	for field-effect transistor applications \cite{zhao2021quantum, sun2021performance, nandan2021two}.
	Beyond pristine monolayer, \ce{MoSi2N4} has been predicted to be half metallic with N or Si vacancies
	\cite{ray2021inducing} and excellent optical absorption in visible light range 
	\cite{jian2021strained} among many other properties, which provide many potential applications in electronics and photonics.

	\begin{figure*}[pt]
		\centering
		\includegraphics[width=1\textwidth]{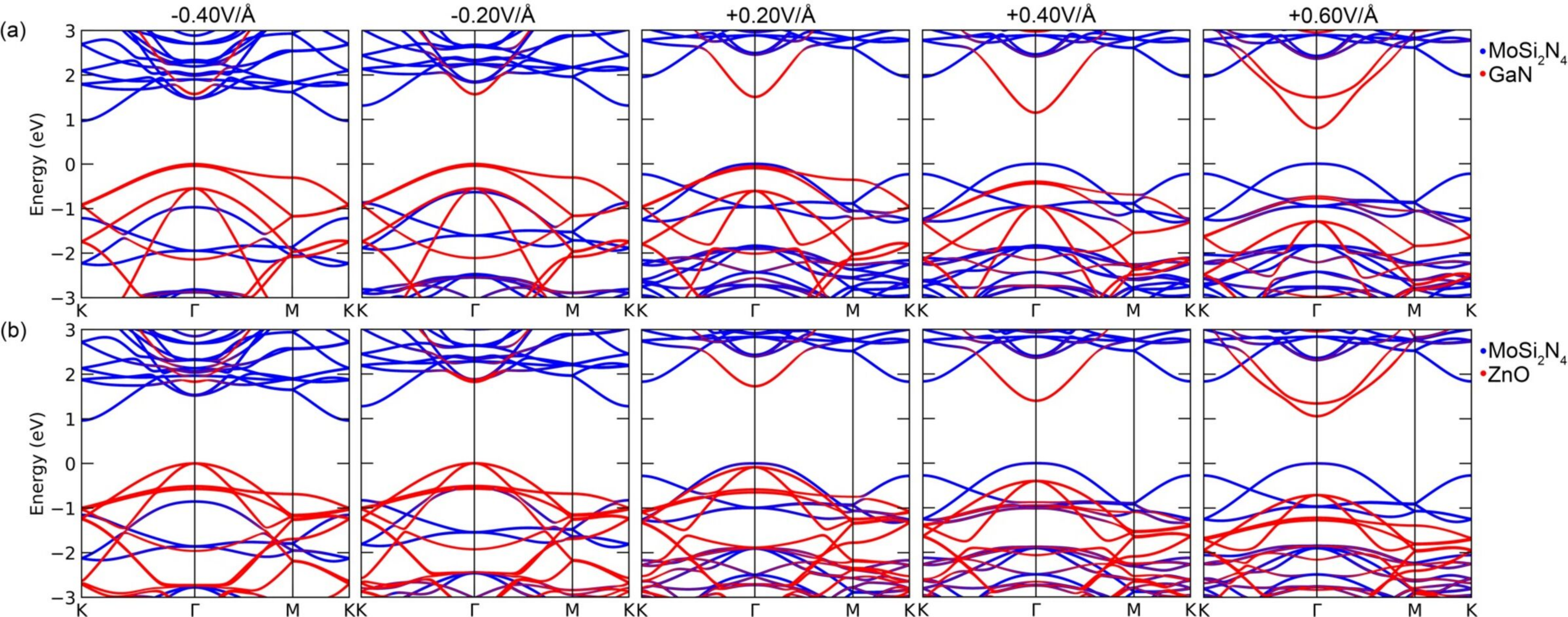}
		\caption{(a) Band structure of \ce{MoSi2N4}/GaN when subjected to external electric fields, with corresponding external electric field values at the top. (b) Band structure of \ce{MoSi2N4}/ZnO when subjected to external electric fields, with corresponding external electric field values at the top.}%
		\label{Figure2}%
	\end{figure*}
	
	\ce{MA2Z4}-based VDWHs is an emerging area, recent studies include \ce{MoSi2N4/TMDC} \cite{cai2021two,bafekry2021van} and 
	Janus \ce{MA2Z4}-based VDWHs \cite{nguyen2021two, zhao2021spin}. 
	Unusual properties, such as electric-field and 
	strain tunable Ohmic-to-contact transition, semiconductor-to-metal transition \cite{wu2021semiconductor}, 
	Type-I-to-Type-II band alignment transition and excellent optical absorption around the visible light regime 
	\cite{nguyen2021two}, have been predicted in various \ce{MA2Z4}-based VDWHs. 
	Nonetheless, the synergy of \ce{MA2Z4} monolayer with wide band gap (WBG) semiconductors, 
	such as GaN and ZnO honeycomb monolayer remains largely unexplored thus far.
	Due to quantum confinement, WBG monolayers are expected to exhibit a larger band gap than the bulk counterparts \cite{al2016two, lee2016tunable}. WBG semiconductors have been widely explored for optical \cite{rong2016high, tian2020research, ben20212d} 
	and electronic device applications \cite{mahmud2018high, apl}. 
	WBG semiconductor typically have a band gap of $> 2$ eV \cite{yoshikawa2007development}, making them excellent for optoelectronic applications \cite{YUAN2021100549} due to higher excitonic binding energy that is robust
	against thermal fluctuations.
	The wider band gap is also beneficial for devices operating at higher temperatures and
	higher voltage operation \cite{fletcher2017survey, zhang2015new, ding2019review, jagadish2011zinc}. 
	In particular, 2D ZnO has been experimentally synthesised \cite{hong2017atomic} and investigated for photocatalytic \cite{gang2021facile} and photodetector capabilities \cite{al2012uv}. On the other hand, low-cost and easy growing of high quality, transferable 2D GaN is challenging due to the parent compound being a non-layered, non-exfoliatable compound. Thus, ongoing research dominantly focuses on the growth of high-quality GaN monolayers \cite{al2016two,chen2018growth}. Nevertheless, novel applications, such as flexible electronics \cite{glavin2017flexible}, light-emitting diodes \cite{chung2010transferable} and piezoelectric strain-gated logic gates \cite{yu2013gan}, have been demonstrated recently, thus suggesting the potential of 2D GaN in electronics and optoelectronics applications. Beyond GaN, other 2D WBG semiconductors, such as GaSe, GaS and SnS$_2$, have been also been fabricated. These monolayers exhibit fast response times and high photoresponsivity \cite{hu2012synthesis,hu2013highly}, good compatibility in flexible device applications \cite{hu2013highly}, and can be incorporated for logic gates applications with large on-off ratios \cite{song2013high}.
	
	Motivated by the potentials of \ce{MA2Z4} based WBG semiconductors, we perform first-principle calculations 
	on the electronic and structural properties of \ce{MoSi2N4}/GaN and \ce{MoSi2N4/ZnO} VDWHs by using density functional theory (DFT). 
	We find that \ce{MoSi2N4}/GaN and \ce{MoSi2N4}/ZnO are Type-I direct band gap and Type-II indirect band gap VDWHs, respectively.
	Interestingly, an external electric fields perpendicular to the plane of the VDWHs can be used to drive a transition between direct and indirect band gaps, thus indicating a field-effect tunable optoelectronic properties of \ce{MoSi2N4}/GaN and \ce{MoSi2N4}/ZnO VDWHs.  
	Additionally, the electronic properties, band alignment and the direct/indirect band gap nature of \ce{MoSi2N4}/GaN and \ce{MoSi2N4}/ZnO can be further controlled under
	mechanical compression and strain. 
	These results suggest the potentials of \ce{MoSi2N4}/WBG-2D-semiconductor VDWHs as a versatile material platform for tunable electronic and optoelectronic device applications. 
	
	All simulations are carried out using DFT implemented in the Vienna \textit{Ab initio} Simulation Package 
	\cite{kresse1993ab, kresse1994ab, kresse1996efficiency, kresse1996efficient}. 
	PAW pseudopotentials \cite{kresse1999ultrasoft} are used to simulate the ion electron bonding and the layers are relaxed 
	using GGA PBE \cite{perdew1996generalized} with Grimme DFT-D3 vdW \cite{grimme2010consistent} interactions between the monolayers. It is known that PBE underestimates the band gaps and that HSE06 will produce better band gaps for 2D GaN and ZnO \cite{al2016two,supatutkul2017electronic}. Using PBE allows the effect of strain and external electric fields to be studied without the high computational costs of HSE06 and has been used before in previous literature \cite{chen2020electronic,huang2014tunable}, so PBE results are used and presented here. Crucially, the general trend of the band structure and band alignment are expected to be sufficiently captured by PBE. We thus expect our key findings to be qualitatively accurate using PBE calculations.
	All materials were sampled using a gamma-centred Brillouin zone at 11 $\times$ 11 $\times$ 1 using the 
	Monkhorst-Pack grid \cite{monkhorst1976special}. 
	Ionic force convergence was set to 0.01 V/{\AA} and electronic convergence was set to $10^{-8}$ eV.
	A vacuum layer of 20 \AA was used to prevent interactions between periodic layers.
	The energy cutoff at 500 eV was made to allow comparisons between materials and different electric field and strain settings. 
	Dipole corrections are enabled in all calculations for consistency. Spin orbit coupling was not considered in this paper, since \ce{MoSi2N4} \cite{hong2020chemical}, GaN and ZnO are non-magnetic semiconductors. Thus, spin is unlikely to play a major role in the properties of the studied VDWH, with spin-orbit coupling showing only small differences  \cite{onen2016gan, kasper2020theoretical, saoud2015band, islam2021tunable}. However, we note that the magnetic properties of VDWH can be altered by means of doping \cite{zhao2016tuning}, hydrogen adsorption \cite{gonzalez2019tuning} or proximity effects to magnetic materials \cite{wu2022giant}, thus suggesting further avenues for magnetic and spintronic properties engineering of MoSi$_2$N$_4$-based VDWH.
	
	\begin{figure*}[ptb]
		\centering
		\includegraphics[width=1\textwidth]{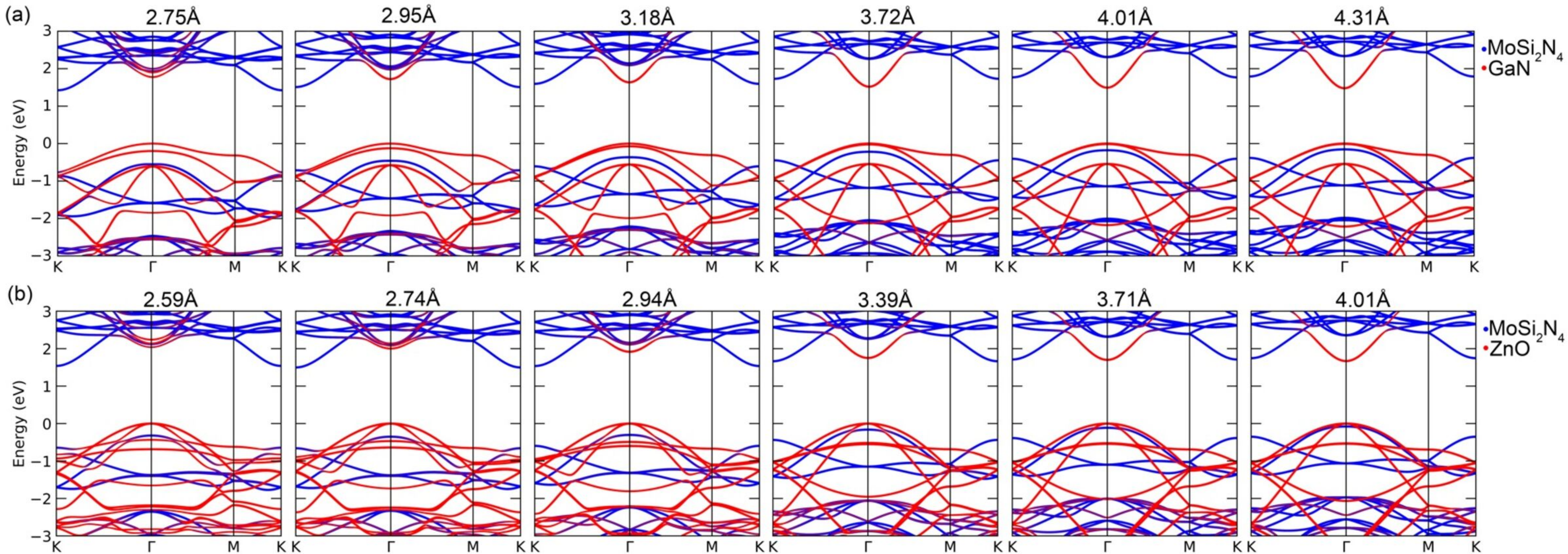}
		\caption{(a) Band structure of \ce{MoSi2N4}/GaN, with corresponding interlayer distance at the top. 
			(b) Band structure of \ce{MoSi2N4}/ZnO, with corresponding interlayer distance at the top}%
		\label{Figure3}%
	\end{figure*}
	
	\ce{MoSi2N4}, GaN and ZnO layers are constructed using previously reported experimental and computational lattice parameters \cite{ray2021inducing, hong2020chemical, ben20212d, al2016two}. 
	The bond lengths of the fully relaxed monolayers are 2.087{\AA} for Mo-N, 1.858 {\AA} for Ga-N and 1.906 {\AA} for Zn-O, 
	which are in agreement with those previously reported. 
	Several stacking configurations were built using QuantumATK, with strain distributed among the GaN and ZnO layer. The binding energy is calculated as follows: 
	$E_b=(E_{vdW}-E_{\ce{MoSi2N4}}-E_{\mu})/34$, where $\mu$ is GaN or ZnO. 
	The binding energy of all configurations was found to be negative, 
	suggesting their energetic stability. The binding energy was all VDWHs very similar to within 0.0001 eV per atom, hence the slightly lower energy was chosen. The final stacking configuration has a binding energy of -0.0158 eV/atom for \ce{MoSi2N4}/GaN 
	and -0.0202 eV/atom for \ce{MoSi2N4}/ZnO. The overall lattice strain for \ce{MoSi2N4}/GaN is 2.68 \% and 0.92 \% for \ce{MoSi2N4}/ZnO.
	
	The chosen VDWHs consist of 2 $\times$ 2 unit cells of \ce{MoSi2N4} 
	and $\sqrt{3} \times \sqrt{3} $ unit cells of GaN or ZnO. The 2 $\times$ 2 supercells are shown in Figs. 1(a) and (b). The fully relaxed \ce{MoSi2N4}/GaN
	and \ce{MoSi2N4}/ZnO have an interlayer distance of 3.44 {\AA} and 3.15 {\AA}, respectively, which are well above the vdW radius of 
	2.05 {\AA} (gallium) \cite{batsanov2001van}, 1.55 {\AA} (nitrogen) \cite{batsanov2001van} and 1.52 {\AA} (oxygen) \cite{batsanov2001van}. 
	The electronic band structures of the isolated \ce{MoSi2N4}, GaN and ZnO monolayers exhibit the band gaps of 1.796 eV, 2.118 eV, and 1.648 eV, respectively.
	The work function is calculated for the monolayers and VDWHs as $W=E_{vacuum}-E_{Fermi}$, where $E_{vacuum}$ is the vacuum energy and $E_{Fermi}$ is the Fermi level. 
	The calculated $E_{vacuum}$ are 4.6 eV, 0.91 eV and 0.54 eV for \ce{MoSi2N4}, GaN and ZnO
	respectively. Correspondingly, the work functions of \ce{MoSi2N4}, GaN and ZnO are calculated as 5.25 eV, 5.23 eV and 5.36 eV, respectively. 
	
	The electronic band structures of the modified monolayers based on the structural parameters taken from the VDWHs are plotted in in Supplementary Materials Figure S1.
	Compared to the free-standing monolayers, the band gaps are modified to 1.952 eV, 1.482 eV and 1.648 eV for \ce{MoSi2N4}, GaN and ZnO, respectively. 
	The overall \ce{MoSi2N4}, GaN and ZnO band structures are preserved 
	in forming the VDWHs. 
	However, the conduction band of GaN and ZnO at $\Gamma$ is split from a single continuous band into avoided crossings across multiple \ce{MoSi2N4} conduction bands. The \ce{MoSi2N4}/GaN has a Type-I direct band gap of 1.56 eV at $\Gamma$-point, with both VBM and CBM contributed by the GaN.
	In contrast, \ce{MoSi2N4}/ZnO has a Type-II indirect band gap of 1.60 eV between the VBM at the $\Gamma$-point as contributed by ZnO, and the CBM at the $K$-point as contributed by \ce{MoSi2N4} CBM. 
	
	We now examine the electron transfer in Figs. 1(e) and 1(f) for \ce{MoSi2N4}/GaN and \ce{MoSi2N4}/ZnO, respectively.
	In general, \ce{MoSi2N4}/GaN has a nett transfer of electrons from \ce{MoSi2N4} to GaN, while \ce{MoSi2N4}/ZnO has a nett transfer of electrons in the opposite direction from ZnO towards \ce{MoSi2N4}.
	In Fig. 1(e), the electron distributions of the \ce{MoSi2N4}/GaN are shown at the cross-sectional and the top views. The electrons accumulate at the contact interface and generally spread out along the interface.  
	In contrast, the electron charge accumulation and depletion in concentric shells around each atom in ZnO for the \ce{MoSi2N4}/ZnO as shown in Fig. 1(f).
	The differential charge charge densities are calculated via
	$\Delta \rho=\rho_{vdW}-\rho_{\ce{MoSi2N4}}-\rho_{\mu}$, where $\mu$ is GaN or ZnO, $\rho_{vdw}$ is the charge density of the VDWH and $\rho_{\mu}$ is the charge density of the individual monolayer.  
	Comparing the differential charge density plots in Figs. 1(e) and (f), it is seen that the peak magnitude of the charge transfer in \ce{MoSi2N4}/GaN is 40 times smaller than that of \ce{MoSi2N4}/ZnO, due to the smaller work function difference between GaN and \ce{MoSi2N4} versus that of ZnO and \ce{MoSi2N4}. 
	The sharp peaks in the differential charge density plot also reveals a significant electron redistribution within the ZnO monolayer but not the \ce{MoSi2N4}, which is in strong contrast to the case of \ce{MoSi2N4}/ZnO VDWH where the electron distributions occur with comparable strength in both \ce{MoSi2N4} and GaN. 
	
	We next examine the heterostructures response to external electric fields. An external electric field was added in a self-consistent manner on relaxed heterostructures in steps of 0.2V/{\AA} in Fig. 2.
	To illustrate the effect of an external electric field on the VDWHs, the differences in charge density are calculated as: $\Delta \rho (E) =\rho(E)-\rho_{\text{initial}}$ where $E$ is the external electric field, $\rho(E)$ and $\rho_{\text{initial}}$ are the charge density of the VDWHs with and without electric field, respectively. 
	The external electric field induces electron depletion on GaN or ZnO, and electron accumulation on \ce{MoSi2N4} due to the electrostatic-induced charge redistributions (see Supplementary Materials for the charge density difference and plane-averaged electrostatic potential plots).
	Thus, at increasing positive electric field the energy of \ce{MoSi2N4} bands are raised relative to GaN and ZnO, while the GaN and ZnO bands are lowered relative to \ce{ MoSi2N4} [see Figs. 2(a) and 2(b) for \ce{MoSi2N4}/GaN and \ce{MoSi2N4}/ZnO, respectively]. 
	Correspondingly, the CBM of both GaN and ZnO at the $\Gamma$-point decreases in energy. Both VDWHs thus undergo a transformation into 
	the Type-II direct band gap VDWHs, featuring the CBM from GaN and ZnO and VBM from \ce{MoSi2N4}. 
	
	
	\begin{figure*}[ptb]
		\centering
		\includegraphics[width=0.9\textwidth]{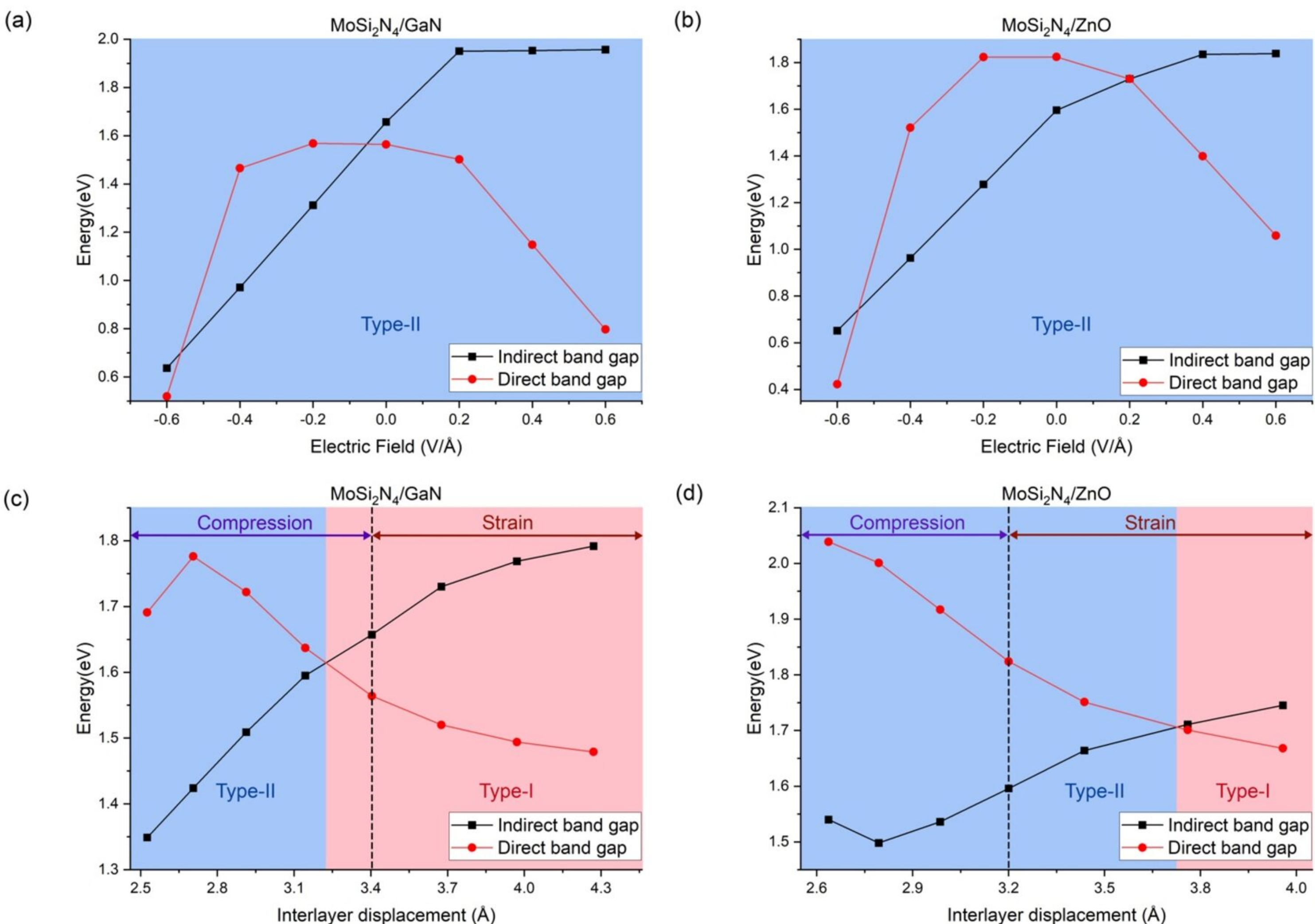}
		\caption{(a) Band gap plots of \ce{MoSi2N4}/GaN under external electric field. 
			(b) Band gap plots of \ce{MoSi2N4}/ZnO under external electric field. 
			(c) Band gap plots of \ce{MoSi2N4}/GaN under strain and compression. 
			(d) Band gap plots of \ce{MoSi2N4}/ZnO under strain and compression}%
		\label{Figure4}%
	\end{figure*}
	
	Under a negative electric field, electron accumulation is induced on GaN and ZnO heterostructure interface and depleted on 
	\ce{MoSi2N4}. 
	In this case, the GaN and ZnO energy is raised relative to \ce{MoSi2N4}, shifting the band structure of GaN and ZnO upwards generally. 
	This results in the CBM of \ce{MoSi2N4} at $K$-point being lower in energy than the $\Gamma$-point CBM of GaN and ZnO, 
	turning both VDWHs into an indirect band gap Type-II VDWHs [see Figs. 2(a) and 2(b) for \ce{MoSi2N4}/GaN and \ce{MoSi2N4}/ZnO, respectively]. 
	

	Experimentally, the interlayer distance can be tuned by several methods during material fabrication stage, such as insertion of hBN buffer layer \cite{fang2014strong} or via nanomechanical pressure \cite{dienwiebel2004superlubricity}. We thus investigate how mechanical strain or compression applied perpendicularly to the VDWHs influences the electronic band structures by changing the interlayer spacing in steps of 0.3 {\AA}.  [see Figs. 3(a) and 3(b) for \ce{MoSi2N4}/GaN and \ce{MoSi2N4}/ZnO, respectively]. 
	%
	For compression, we decrease the bond lengths of Mo-N from 2.080{\AA} and 2.088{\AA} from the initial \ce{MoSi2N4}/GaN and \ce{MoSi2N4}/ZnO VDWHs, respectively, to 2.062{\AA} and 2.068{\AA} in \ce{MoSi2N4}/GaN and \ce{MoSi2N4}/ZnO VDWHs, respectively. 
	%
	When the VDWHs are compressed [see Figs. 3(a) and (b)], \ce{MoSi2N4} is observed to decrease in energy with respect to GaN and ZnO in general. 
	These changes are similar to those observed when the VDWHs are subject 
	to an external electric field in the positive $z$-direction. 
	Furthermore, the width of the vacuum tunneling potential barrier between the monolayers decreases under compression (see the plane-averaged electrostatic potential profile plots in the SM), which leads to easier tunneling of electrons from GaN and ZnO to \ce{MoSi2N4}. 
	The GaN or ZnO sublayer increases in energy with increasing compression, with bands shifting higher relative to \ce{MoSi2N4}. 
	For \ce{MoSi2N4}/GaN, the CBM of GaN at the $\Gamma$-point retracts into the higher energies, resulting in the transformation from direct band gap Type-I into indirect band gap Type-II band alignment. 
	For \ce{MoSi2N4}/ZnO, an indirect band gap Type-II band alignment is similarly observed at large compression. 
	
	For tensile strain, the bond lengths of Mo-N increases from 2.080{\AA} and 2.088{\AA} in initial \ce{MoSi2N4}/GaN and \ce{MoSi2N4}/ZnO VDWH to 2.083{\AA} and 2.090{\AA} in the strained VDWHs at maximum strain. 
	GaN and ZnO is observed to decrease in energy with the band structure shifting downwards relatively to \ce{MoSi2N4}. 
	However, the increasing width of the potential barrier between 
	the monolayers as the interlayer distance is increased (see SM) severely limits the ability of electrons to tunnel across and reduces the coupling between the two materials. 
	Thus, the changes in the band structure less dramatic as compared to the case of compression. 
	For \ce{MoSi2N4}/GaN, the band structure remains a direct band gap Type-I band alignment under increasing strain with decreasing change in direct band gap energy. 
	For \ce{MoSi2N4}/ZnO, the band structure changes from indirect band gap Type-II to direct band gap Type-I band alignment
	as the ZnO CBM energy decreases relative to \ce{MoSi2N4} CBM. 
	
	The properties of the band gap and the heterostructure types are summarized in Fig. 4, with blue-shaded regions representing Type-II alignment and 
	red-shaded regions representing Type-I alignment. 
	The VDWHs retain the Type-II band alignment 
	under external electric fields in both directions as shown in Figs. 4(a) and 4(b) for \ce{MoSi2N4}/GaN and \ce{MoSi2N4}/ZnO, respectively. 
	However, the heterostructures changes between direct and indirect band gap when the electric fields are varied in the range of 0.0 V/{\AA} to -0.6 V/{\AA} for \ce{MoSi2N4}/GaN and +0.2V/{\AA} to -0.6 V/{\AA} for \ce{MoSi2N4}/ZnO. 
	This is in contrasts to the case of compression and strain, as shown in Figs. 4(c) and 4(d) for \ce{MoSi2N4}/GaN and \ce{MoSi2N4}/ZnO, respectively.
	Under compression, \ce{MoSi2N4}/GaN changes from 
	direct band gap Type-I to indirect band gap Type-II  band alignment, while \ce{MoSi2N4}/ZnO changes from indirect band gap Type-I to indirect band gap Type-II band alignments.
	Under strain, \ce{MoSi2N4}/GaN remains unchanged while \ce{MoSi2N4}/ZnO changes from Type-I indirect band gap to Type-I direct band gap 
	under strain. 
	
	In summary, the electronic properties of \ce{MoSi2N4}-based wide band gap semiconductor van der Waals heterostructures are studied using density functional theory simulations. 
	The \ce{MoSi2N4}/GaN and \ce{MoSi2N4}/ZnO heterostructures are in direct band gap Type-I and indirect band gap Type-II band alignment, respectively.
	Differential charge density analysis suggests that the electron redistribution upon forming the contact heterostructure is much stronger in \ce{MoSi2N4}/ZnO than that in \ce{MoSi2N4}/GaN. 
	Furthermore, we show that the application of an external perpendicular electric field or mechanical strain effectively alter the band gap type and the band alignment type of the heterostructures, thus offering a tuning knob for engineering the electronic and optoelectronic properties of the heterostructures. 
	These findings reveal the potentials of \ce{MoSi2N4}-based wide band gap van der Waals heterostructures as a versatile platform for designing novel tunable optoelectronics.
	
	\section*{Supplementary Material}
	
	See supplementary material for the electronic band structures of isolated monolayers, and the plane-averaged differential charge and electrostatic potential profiles of the MoSi$_2$N$_4$/ZnO and MoSi$_2$N$_4$/GaN heterostructures.
	
	\section*{ACKNOWLEDGMENTS}
	
	This work is supported by A*STAR AME IRG (A2083c0057) and Singapore University of Technology and Design Start-Up Research grant (Project No. SRG SCI 2021 163). J.Q.N. acknowledge the PhD Scholarship support from the Singapore University of Technology and Design. The calculations were carried out using the computational resources provided by the Titan supercomputing facility in SUTD and the National Supercomputing Centre (NSCC) Singapore. 
	
	\section*{Author Declarations}
	
	\subsection*{Conflict of Interest}
	\noindent The authors declare that there are no conflicts of interest.
	
	\subsection*{Author Contributions}
	\noindent J.Q.N. performed the simulations and data analysis. Q.W., L.K.A. and Y.S.A. supervised the project. All authors contributed to the writing and the revision to this work.
	
	\section*{Data Availability} 
	
	The data that support the findings of this study are available from the corresponding author upon reasonable request.

	\bibliographystyle{apsrev}
	
	\providecommand{\noopsort}[1]{}\providecommand{\singleletter}[1]{#1}%

\end{document}